\documentclass[showpacs,superscriptaddress,twocolumn]{revtex4-1}%
\usepackage{hyperref}
\usepackage{graphicx}
\usepackage{amsfonts}
\usepackage{amssymb}
\usepackage{amsmath}

\begin{document}

\title{Interval identification of FMR parameters
for spin reorientation transition in (Ga,Mn)As
}

\author{M.W.~\surname{Gutowski}}
\email[Corresponding author: ]{marek.gutowski@ifpan.edu.pl}
\affiliation{Institute of Physics, Polish Academy of Sciences,
Warszawa, Poland}

\author{W.~\surname{Stefanowicz}}
\affiliation{Institute of Physics, Polish Academy of Sciences,
Warszawa, Poland}

\author{O.~\surname{Proselkov}}
\affiliation{Institute of Physics, Polish Academy of Sciences,
Warszawa, Poland}

\author{J.~\surname{Sadowski}}
\affiliation{Institute of Physics, Polish Academy of Sciences,
Warszawa, Poland}
\affiliation{MAX-lab, Lund University, Lund, Sweden}

\author{M.~\surname{Sawicki}}
\affiliation{Institute of Physics, Polish Academy of Sciences,
Warszawa, Poland}

\author{R.~\surname{\.Zuberek}}
\affiliation{Institute of Physics, Polish Academy of Sciences,
Warszawa, Poland}
%
%
\begin{abstract}
In this work we reports results of ferromagnetic resonance
studies of a 6\% $15\,$nm (Ga,Mn)As\ layer, deposited on
$(001)$-oriented GaAs. The measurements were performed with
in-plane oriented magnetic field, in the temperature range
between $5\,$K and $120\,$K.\ We observe a~temperature induced
reorientation of the effective \mbox{in-plane} easy axis from
$[\overline{1}10]$ to $[110]$ direction close to the Curie
temperature.\ The behavior of magnetization is described by
anisotropy fields,\ $H_{\rm eff}\,(=4\pi{M}-H_{2\perp})$,
$H_{2\parallel}$, and $H_{4\parallel}$.\ In~order to precisely
investigate this reorientation, numerical values of anisotropy
fields have been determined using powerful -- but still largely
unknown -- interval calculations.\ In simulation mode this
approach makes possible to find \emph{all} the resonance fields
for arbitrarily oriented sample, which is generally intractable
analytically.\ In~`fitting' mode we effectively utilize full
experimental information, not only those measurements performed
in~special, distinguished directions, to~reliably estimate the
values of~important physical parameters as well as their
uncertainties and correlations.
\end{abstract}


\maketitle

\section{Motivation}

Despite numerous and intensive studies, an origin of the in-plane uniaxial
magnetic anisotropy in (Ga,Mn)As remains unknown.\ However, both its
strength and orientation can be described on the ground of \textit{p-d} Zener
model assuming the existence of a~fictitious epitaxial strain
\mbox{\cite{Sawicki:2005_PRB,Zemen:2009_PRB,Stefanowicz:2010_PRB}}.\
On the other hand, as this is the leading magnetic anisotropy at elevated
temperatures and that the means of its control have already been
demonstrated \cite{Chiba:2008_N}, further studies on this intriguing
property are timely and important.\  In particular, a~presence of the
temperature-induced $90^{\circ}$ rotation of the direction of the easy axis
\cite{Sawicki:2005_PRB} may ease the vector manipulation of magnetization in
future devices.  In this communication we report on the technical analysis
and results of FMR studies of such a~(Ga,Mn)As layer that exhibit the easy axis
rotation at temperatures close to its Curie temperature.

The free energy density for our system, expressed by anisotropy fields
($H$'s) rather than by more customary anisotropy constants ($K$'s), has
the form ($g\mu_{\textrm B}HM$ (Zeeman), and inactive out-of-plane fourfold
anisotropy term have been omitted):
\begin{eqnarray}
F &=& 2\,\pi\,M^{2}\cos^{2}\theta \label{shape}\\
&& -\frac{1}{2}\,M\,H_{2\perp}\,\cos^{2}\theta \label{u-out}\\
&& -\frac{1}{2}\,M\,H_{2\parallel}\,\sin^{2}\theta\,\sin^{2}\left(\varphi-\frac{\pi}{4}\right) \label{u-in}\\
&& -\frac{1}{16}\,M\,H_{4\parallel}\,\left(3+\cos 4\varphi\right)\,\sin^{4}\theta \label{4-in}
\end{eqnarray}
where (\ref{shape}) is shape anisotropy, (\ref{u-out}) -- ordinary
uniaxial out-of-plane anisotropy, (\ref{u-in}) -- uniaxial in-plane
anisotropy, 
and (\ref{4-in}) -- fourfold, in-plane anisotropy.

Here polar angles angles $\theta$\ and\ $\varphi$\ refer to the
orientation of the magnetization vector, $\vec{M}$, \textbf{not} to the
orientation of an external field $\vec{H}$.

\begin{figure}[h]
    \includegraphics[keepaspectratio,angle=-90,width=\hsize]{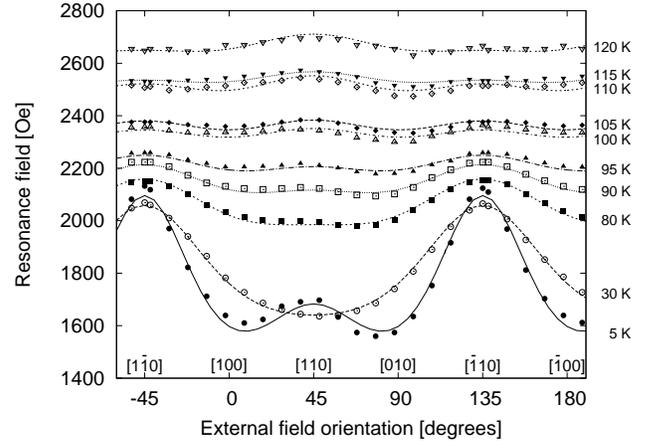}
    \caption{Experimental values of FMR fields \emph{vs.} external field orientation
    and temperature (points).\ Lines show computed spectra.
    }\label{meas}
\end{figure}

Original experimental data, taken at fixed frequency $9.378\,$GHz, are
shown in Fig.~\ref{meas}, together with simulated spectra.

The numerical values of three parameters:\
$H_{\textrm{eff}}=4\pi M - H_{2\perp}$,\ (where
$H_{2\perp}\,\equiv\,2K_{2\perp}/M$),\ ~$H_{2\parallel}$\ ~and\
~$H_{4\parallel}$
were determined
for each temperature separately.

\section{Outline of numerical procedure}
We are using interval calculations not only for accurate simulations of
FMR spectra when the values of all relevant parameters are known, but
also to estimate (`fit') such parameters, together with their
uncertainties and correlations, utilizing full information hidden
in~experimental data.\ 
An excellent
introduction to interval arithmetics
can be found 
in~\cite{site}.\ Here we limit ourselves to the very brief description:
{\bf i})~an interval $\mathbf{x}$ is a~bounded set of real numbers:
$\mathbf{x}\equiv[a,b] = \left\{\mathbb{R}\ni x: a\leqslant x\leqslant
b\right\}$,\ {\bf ii})~it is possible to perform arithmetic operations,
evaluate functions, etc., using intervals instead of numbers, {\bf iii})~the so
obtained results are intervals as well, {\bf iv})~the multidimensional
intervals are called boxes for obvious reasons, and {\bf v})~ordinary real
numbers may be identified with ('thin') intervals, for example $7=[7,7]$.\
Similarly, $f(\mathbf{x})$ is an interval containing \emph{all} the
possible results of evaluation of $f(x)$ when $x\in\mathbf{x}$.\
Unfortunately, $f(\mathbf{x})$ usually overestimates the range of true
values of $f(x)$ -- but always contains them all.

The unconventionality of our approach to fixed frequency FMR data fitting
is that we try to adjust unknown parameters in such a way that the
classical formula
\begin{equation}
\left(\frac{\omega}{g\mu_{\mathrm B}}\right)^{2}=
\frac{1}{(M\sin\theta)^{2}} \left[\frac{\partial^{2}F}{\partial\theta^{2}}
\cdot\frac{\partial^{2}F}{\partial\varphi^{2}}
-\left(\frac{\partial^{2}F}{\partial\theta\;\partial\varphi}\right)^{2}\right]
\label{classic}
\end{equation}
is satisfied for each experimental datum.

The main difficulty lies in the fact that the partial derivatives of
free energy density have to be evaluated at (stable) equilibrium
position of magnetization vector, characterized by two (initially
unknown) polar angles $\varphi$ and $\theta$, and being the solution(s)
of the equation's system
\begin{equation}
\nabla F = (\partial{F}/\partial{\theta},
\partial{F}/\partial{\varphi}) = 0.
\end{equation} 
This makes the inversion of formula (\ref{classic}), to obtain
resonance field(s) as a~function of microwave frequency $\omega$,
essentially impossible.

\medskip
Our algorithm operates on a~list $\mathcal{L}$ of boxes.\ At the
beginning, the list contains only one member -- the initial search
domain.\ Further we perform following steps:
\begin{enumerate}
\item select the biggest box from the list $\mathcal{L}$
\item bisect its longest edge obtaining two offspring boxes,
then remove the parent box from the list~$\mathcal{L}$,
\item investigate each one of the two offspring and:\\
-- discard it, if infeasible, or\\
-- put it back on the list $\mathcal{L}$, otherwise.
\end{enumerate}
The above procedure is repeated until the list contains only small
boxes.\ The rest is easy, provided the final cluster of boxes consists
of only one connected component.\ For details on how to calculate mean
values, variances and correlations between the searched parameters see
\cite{Zawoja}.\ Here we only clarify when the box is considered
infeasible.\ First of all, each box is characterized by it's
\emph{quality factor}, computed as an interval quantity
$\mathbf{Q}=[\underline{Q}, \overline{Q}]$.\ For each resonance field
(and its corresponding orientation) we try to evaluate the interval
value of r.h.s. of expression (\ref{classic}).\ If the upper bound of
so calculated interval is negative, then the currently considered box
of searched parameters is infeasible as it cannot describe any
resonance at all.\ If all experimental data pass this test, then the
ranges of their resonance frequencies can be found from (\ref{classic})
and the interval $\mathbf{Q}$ can be evaluated.\ Again, for each data
element $i$ the absolute difference between both sides of
(\ref{classic}) is calculated, producing the interval
$\mathbf{\Delta}\mathbf{\omega}_{i}$.\ 
Finally,
$\mathbf{Q}=\max_{i}\mathbf{\Delta}\mathbf{\omega}_{i}$ (other choices
are possible but we prefer this one).\ The box is considered infeasible
when it's $\underline{Q}$ exceeds the reference value, which is equal
to the lowest $\overline{Q}$ ever seen during calculations.

It remains to comment on resonance frequency calculation.\ The angles
$(\theta, \varphi)$ are unknown and should be determined prior to
resonance frequency evaluation, for each data element separately.\
Starting from full ignorance ($\theta\in[0,\pi]$,
$\varphi\in[0,2\pi]$), we divide this initial 2D box until its edges
are shorter than, say, $0.5^{\circ}$.\ Discarded are all boxes
satisfying either of two conditions:
$0\notin\partial{F}/\partial\theta$ or
$0\notin\partial{F}/\partial\varphi$ -- they certainly cannot contain
the equilibrium position of the vector $\vec{M}$.\ Needles to say that
failure to find such a~position immediately invalidates the searched
parameter box.

\section{Results and discussion}
\begin{figure}[h]
    \includegraphics[keepaspectratio,angle=-90,width=\hsize]{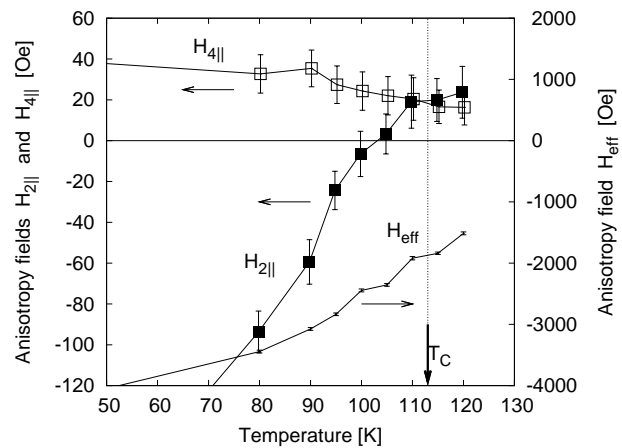}
    \caption{Anisotropy fields \emph{vs.} temperature -- computed
    values.\ Lines are eye-guides only.\ Apparent lack of smoothness
    may be attributed to inaccurate temperature readings (the
    measurements were taken during two separate sessions).\ For better
    visibility, the curves $H_{2\parallel}$ and $H_{4\parallel}$ are
    shifted horizontally by $-0.2\,$K and $+0.2\,$K, respectively. 
    }\label{3const}
\end{figure}

\begin{figure}[h]
    \includegraphics[keepaspectratio,angle=-90,width=\hsize]{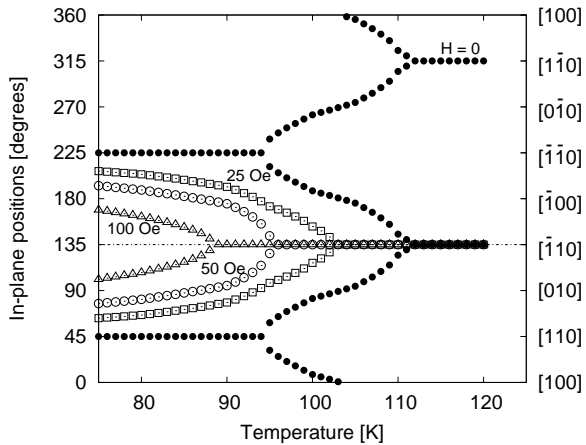}
    \caption{Equilibrium position of magnetization vector \emph{vs.}
    temperature for various strengths of the external field.\ The field
    is~always directed along high-temperature easy axis
    $[\overline{1}10]$, marked also as $135^{\circ}$, and its strength
    is as indicated.  }\label{H135}
\end{figure}

Following the procedure described in previous section, we have computed
the values of parameters determining the free energy density
(\ref{shape}--\ref{4-in}).\ During computation $g$ was kept fixed at
$2.00$.\ The results are presented in Fig.~\ref{3const}.\ The numerical
values are consistent with those reported by others
\cite{Furdyna-JPCM,Furdyna-x,Furdyna1}, obtained by the same technique
but at lower temperatures.
Using those results, without any prior smoothing, we were able to
determine the directions of spontaneous magnetization in the
interesting temperature range, near $T_{\mathrm{C}}$.\ In
Fig.~\ref{H135} one can see that, in absence of the external field,
there are only two such directions, antiparallel to each other, not
four as one might expect from symmetry arguments.\ The presence of
external magnetic field, oriented along $[\overline{1}10]$ (Earth's
field is sufficient) breaks even this symmetry, and the spontaneous
magnetization aligns itself exactly along the external field above the
transition temperature.\ Below the transition temperature the
spontaneous magnetization deviates considerably from its 'natural'
$[110]$ (or, equivalently, $[\overline{1}\overline{1}0]$) position.
This means that precise examination of spontaneous magnetization,
especially of its components perpendicular to the external field,
is difficult.\ The temperature range, in which the reorientation occurs,
is also very sensitive to the presence of even very weak field.

The behavior of anisotropy fields near $T_{\mathrm{C}}$, estimated as
$\sim\!113\;$K for our sample, is somewhat
unusual: they all should go to $0$ as $T\rightarrow\,T_{\mathrm{C}}$ --
but they don't.\ In addition, the simulated angular FMR dependencies
(Fig.~\ref{meas}) seem not so accurate as one might expect.\ Those
intriguing facts, together with high reliability of interval analysis,
strongly suggest that the uniaxial in-plane anisotropy may be
incorrectly accounted for in\ (\ref{shape}--\ref{4-in}).\ Since the
samples exhibiting this special behavior are very thin, in nm range,
then it is quite possible that the source of uniaxial anisotropy may be
related to surface effects.\ Indeed, AFM surface studies of MBE-grown
samples \cite{ripples} revealed the presence of well ordered ripples,
parallel to $[1\overline{1}0]$ direction.\ If this was true, then the
free energy expression should be appended with the surface term \cite{MWG},
proportional to\ $\left|\cos(\varphi-\pi/4)\right|$.\
This, however, is beyond the scope of a~current paper.
\medskip

\section{Conclusions}
The interval calculus has been demonstrated to be a~very powerful
tool for difficult problems of experimental data analysis.\
In particular, it is probably the only method able to utilize
complete experimental information acquired during FMR measurements.\
Its unrivaled reliability
allows us to state a~sound hypothesis concerning the
nature of the somewhat mysterious in-plane uniaxial magnetic anisotropy
observed in thin layers of (Ga,Mn)As grown on $(001)$-oriented GaAs
substrate.

We have also shown that FMR technique is very helpful in accurate
tracing the temperature dependency of magnetic anisotropy in close
vicinity of the Curie temperature, that is where the magnetometric
data are least reliable.\ 


\section*{Acknowledgments}
This work was supported in part by EC Network SemiSpinNet
(PITN-GA-2008-215368) and Polish MNiSW 2048/B/H03/2008/34 grant.

\end{document}